\documentclass{article}
\usepackage{amsfonts}
\usepackage{amssymb}
\usepackage{amsmath}
\usepackage{amsthm}
\usepackage{mathtools}
\begin{document}

\title{Geometry of teleparallel theories}

\author{Alexey Golovnev\\
{\small {\it Centre for Theoretical Physics, the British University in Egypt,}}\\ 
{\small {\it BUE 11837, El Sherouk City, Cairo Governorate, Egypt}}\\
{\small agolovnev@yandex.ru}
}
\date{}

\maketitle

\begin{abstract}

I give a brief introduction to and explain the geometry of teleparallel models of modified gravity. In particular I explain why, in my opinion, the covariantised approaches are not needed and the Weitzenb{\"o}ck connection is the most natural representation of the parallel transport structure. An interesting point is that it also applies to the symmetric teleparallel case. I also share my thoughts on why the teleparallel framework does not seem to be a next rung in the ladder of understanding the real worlds' gravity. At the same time, these theories do have a clear and justified academic interest to them. 

This is a talk presented at the International Conference on Particle Physics and Cosmology dedicated to memory of Valery Rubakov, Yerevan, Armenia, October 2023.

\end{abstract}

\section{The realm of modifying gravity}

Gravity is very successfully described by the famous theory of Albert Einstein. It is one of the best and most beautiful theories we have. Still, we are stubbornly trying to modify it. Why is that? Well, there are plenty of different possible reasons, depending on who you ask. There are mysteries in cosmology. What are the Dark Sectors? Was there inflation, and if yes then how? And if the issues such as $H_0$ tension are real, what are we making out of that? On top of this, there are singularities, inherent and unavoidable. They are mostly hidden whenever one can imagine. But don't we want to have a better understanding of what is going on? And let alone the puzzle of quantum gravity, together with our pathological belief in the mathematically horrendous quantum field theory framework.

Given all the intriguing motivations above, the amazing news we get is that it is extremely difficult to meaningfully modify the theory of General Relativity. We have a huge variety of modified gravity models, maybe even too many of them \cite{reflections}.  But how far have we really gone after all? Simple models such as $f(R)$ are almost nothing new, and can be reformulated as an extra universal force mediated by a scalar field on top of the usual gravity. Attempts at more profound modifications require exquisite care to not encounter ghosts, or other bad instabilities, or total lack of well-posedness, or no reasonable cosmology available, or.... You name it!

Having got the miserable lack of an undoubtful success, it certainly makes sense to try whatever crazy new idea one can think of. And let it lead us to a better understanding of what is a theory of gravity. One possible alternative approach is to describe gravity in terms of different geometry. The classical teleparallel, or the metric teleparallel  approach proceeds in terms of torsion instead of curvature, while the symmetric teleparallel -- in terms of non-metricity. Their common feature is vanishing curvature, and this is the defining property of any teleparallel model.

For an arbitrary given set of connection coefficients $\Gamma^{\alpha}_{\mu\nu}$, the definitions of the basic tensorial quantities are very well-known:
\begin{equation}
\label{tors}
T^{\alpha}_{\hphantom{\alpha}\mu\nu}\equiv \Gamma^{\alpha}_{\mu\nu}- \Gamma^{\alpha}_{\nu\mu}
\end{equation}
for the torsion and
\begin{equation}
\label{nonmet}
 Q_{\alpha\mu\nu}=\bigtriangledown_{\alpha}g_{\mu\nu}
\end{equation}
for non-metricity on a manifold equipped with a metric $g_{\mu\nu}$. Then it is nothing but a textbook exercise, to see that the connection is always given by a simple correction to the Levi-Civita one:
\begin{multline}
\label{gencon}
\Gamma^{\alpha}_{\mu\nu}=\mathop\Gamma\limits^{(0)}{\vphantom{\Gamma}}^{\alpha}_{\mu\nu}(g)+K^{\alpha}_{\hphantom{\alpha}\mu\nu} +L^{\alpha}_{\hphantom{\alpha}\mu\nu}\equiv
\frac12 g^{\alpha\beta}\left(\partial_{\mu}g_{\nu\beta}+\partial_{\nu}g_{\mu\beta}-\partial_{\beta}g_{\mu\nu}\right)\\
+\frac12\left(T^{\alpha}_{\hphantom{\alpha}\mu\nu}+T^{\hphantom{\nu}\alpha}_{\nu\hphantom{\alpha}\mu}+T^{\hphantom{\mu}\alpha}_{\mu\hphantom{\alpha}\nu}\right)-\frac12\left(Q^{\hphantom{\mu\nu}\alpha}_{\mu\nu}+Q^{\hphantom{\nu\mu}\alpha}_{\nu\mu}-Q^{\alpha}_{\hphantom{\alpha}\mu\nu}\right).
\end{multline}
Note in passing that I ascribe the left lower index of a connection coefficient to the derivative, e.g. $\bigtriangledown_{\mu}T^{\nu}\equiv\partial_{\mu}T^{\nu}+\Gamma^{\nu}_{\mu\alpha}T^{\alpha}$. With non-symmetric connections, it is very important to consistently follow a convention about the position of this index.

I would also like to stress that the teleparallel framework is more than just about a metric. If we write the theory down in terms of a metric only, then there isn't much more than the standard Riemannian geometry to be used. Of course, there could formally be a non-metricity if the connection is Levi-Civita for another metric, say $f(\phi) \cdot g_{\mu\nu}$, with an auxiliary scalar field which might be either a new fundamental entity on top of the metric or an effective description of non-linear realisations of a pure-metric model. This is not something excitingly new for geometry. However, every teleparallel theory does have a very geometric addition to its foundational features: the structure of a unique and unambiguous parallel transport. 

If the parallel transport has got an identically zero curvature tensor, then at least locally, its results do not depend on the chosen path. It means that one can randomly choose a basis in the tangent space of a randomly chosen point on the manifold and parallelly transport it to every other point on it. By this procedure, we do get a basis of covariantly constant vector fields. This is the fundamental teleparallel tetrad defining a Weitzenb{\"o}ck connection. If we assume objective existence of the flat connection, the choice of this tetrad is not arbitrary.

I do not agree with the common opinion that it is necessary to have a locally Lorentz covariant description of teleparallel gravity \cite{KrSa, KrCo}, nor with another frequent opinion that such a description is severely problematic \cite{MaUl}. On one hand, unless in TEGR or STEGR or any other way of just reproducing GR, different tetrads for the same metric are physically different objects corresponding to different parallel transport structures. The wish of enjoying the freedom of choosing the tetrads comes from the habit of working in general relativity where those are nothing but arbitrary bases in the linear spaces either representing observers or just serving as a tool for making the calculations convenient. The teleparallel framework is different. On the other hand, even with a fundamentally preferred tetrad, there is still no problem of rewriting all the same story in terms of some another basis \cite{cov}. Moreover, it can sometimes be very convenient to do so \cite{meLor, geo}.

\section{A brief introduction to teleparallel theories}

As it has been mentioned above, the general idea of teleparallel theories is to have geometry with a parallel transport of zero curvature. I would discuss the two most elementary choices of this sort: purely torsion and purely non-metricity. Any further generalisations can be easily derived.

\subsection{Metric teleparallel}

The quest for TEGR action can start from observing the particular case of the formula (\ref{gencon})  for a metric-compatible (i.e. of zero non-metricity) connection :
\begin{equation*}
\Gamma^{\alpha}_{\mu\nu}=\mathop\Gamma\limits^{(0)}{\vphantom{\Gamma}}^{\alpha}_{\mu\nu}(g)+K^{\alpha}_{\hphantom{\alpha}\mu\nu}
\end{equation*}
where the contortion tensor is defined in terms of the torsion tensor (\ref{tors}) as
\begin{equation}
\label{cont}
K_{\alpha\mu\nu}=\frac12\left(T_{\alpha\mu\nu}+T_{\nu\alpha\mu}+T_{\mu\alpha\nu}\right),
\end{equation}
 antisymmetric in the lateral indices. We usually represent such flat connection in terms of an orthonormal tetrad as
\begin{equation}
\label{Weitz}
\Gamma^{\alpha}_{\mu\nu}=e_n^{\alpha}\partial_{\mu}e^n_{\nu}
\end{equation}
with $e^n_{\mu}$ being the dual basis, or matrix inverse, for the basis of orthonormal vectors $e_n^{\mu}$.

The curvature tensors
\begin{equation}
\label{curv}
R^{\alpha}_{\phantom{\mu}\beta\mu\nu}=\partial_{\mu}\Gamma^{\alpha}_{\nu\beta}-\partial_{\nu}\Gamma^{\alpha}_{\mu\beta}
+\Gamma^{\alpha}_{\mu\rho}\Gamma^{\rho}_{\nu\beta}
-\Gamma^{\alpha}_{\nu\rho}\Gamma^{\rho}_{\mu\beta}
\end{equation}
for the two different connections obviously have a quadratic in $K$ expression in their difference. Then making necessary contractions, such as $R_{\mu\nu}=R^{\alpha}_{\phantom{\alpha}\mu\alpha\nu}$ and $R=g^{\mu\nu}R_{\mu\nu}$, we can come to
\begin{equation}
\label{curvs}
0 = R = \mathop{R}\limits^{(0)}+{\mathbb T}+2\mathop{\bigtriangledown_{\mu}}\limits^{(0)}T^{\mu}
\end{equation}
where 
\begin{equation}
\label{torvec}
T_{\mu} = T^{\alpha}_{\hphantom{\alpha}\mu\alpha}
\end{equation}
is the torsion vector while the torsion scalar
\begin{equation}
\label{torsca}
{\mathbb T} = \frac 12 S_{\alpha\mu\nu}T^{\alpha\mu\nu}
\end{equation}
is given in terms of the superpotential
\begin{equation}
\label{super}
S_{\alpha\mu\nu} = K_{\mu\alpha\nu}+g_{\alpha\mu}T_{\nu}-g_{\alpha\nu}T_{\mu}.
\end{equation}

Due to the basic relation (\ref{curvs}), the Einstein-Hilbert action 
$$S_{\mathrm{EH}}=-\int d^4 x \sqrt{-g}\mathop{R}\limits^{(0)}$$ 
is equivalent to the TEGR one, 
$$S_{\mathrm{TEGR}}=\int d^4 x \| e\| \mathbb T,$$  
because the Lagrangians are the same, up to the surface term 
$${\mathbb B} = 2\mathop{\bigtriangledown_{\mu}}\limits^{(0)}T^{\mu}.$$ 
 Of course, this equivalence disappears when we go to modified gravity, for example the $f(T)$ gravity:
\begin{equation}
\label{fT}
S=\int f(\mathbb T)\cdot \| e\| d^4 x.
\end{equation}

Actually, the work of varying the action (\ref{fT}) can be simplified a lot by using this observation of (non-)equivalence \cite{meLor}. After some little exercise, the equation of motion can be written as
\begin{equation}
\label{emfT}
f^{\prime}\mathop{G_{\mu\nu}}\limits^{(0)}+\frac12 \left(f-f^{\prime}{\mathbb T}\right)g_{\mu\nu}+f^{\prime\prime}S_{\mu\nu\alpha}\partial^{\alpha}{\mathbb T}=\kappa {\cal T}_{\mu\nu}
\end{equation}
with ${\cal T}_{\mu\nu}$ being the energy-momentum tensor of the matter. I assume that it is the standard, symmetric and covariantly-conserved tensor, with no hypermomentum around. This is a very nice (convenient and covariant) form of equations. In particular, we immediately find out that, quite naturally, a non-trivial antisymmetric part of equations (\ref{emfT}) appears if and only if $f^{\prime\prime}\neq 0$, and it takes the form of
$$(S_{\mu\nu\alpha}-S_{\nu\mu\alpha})\partial^{\alpha}{\mathbb T}=0.$$
It can be thought of as related to Lorentzian degrees of freedom. And we also see that solutions with a constant value of $\mathbb T$ are very special and do not go beyond the usual GR, unless we are to study perturbations around them.

So far so good... Then the real problems come \cite{issues}. The number of degrees of freedom is not very well known. To my mind, the main reason for that is a variable rank of the algebra of Poisson brackets of constraints. There are different Hamiltonian claims in the literature \cite{Ham1, Ham2, Ham3}. But, what is for sure, is that there must be at least one extra mode \cite{Mink1, Mink2}.  Still, the trivial Minkowski tetrad 
\begin{equation}
\label{trt}
e^A_{\mu}=\delta^A_{\mu}
\end{equation}
 is obviously a legitimate background for a model with $f(0)=0$ in vacuum\footnote{Of course, at the same time I assume $f^{\prime}(0)\neq 0$ for the usual graviton to be alive. } showing a strong coupling regime for all the extra stuff. 

Indeed, in this case (\ref{trt}) we have 
$${\mathbb T}\propto (\partial \delta e)^2 + {\mathcal O}((\delta e)^3),$$ 
and for the quadratic action we then just take 
$$f(\mathbb T)= f_0 + f_1\mathbb T+ {\mathcal O}(\mathbb T^2) =  f_1\mathbb T + {\mathcal O}((\delta e)^3)$$ 
which means accidental restoration of the full Lorentz symmetry, and therefore linearised GR and nothing more. All the new degrees of freedom have disappeared from the quadratic action. What is much less immediately obvious is that, in what concerns dynamical modes, this strong coupling persists also in cosmology \cite{cosm1, cosm2}.

Therefore, all the standard properties of gravitational waves are there, that is the usual two polarisations, propagation at the speed of light, as well as any other property which can be derived from the linearised GR equations. With all the fun which can be caused by this phrase, I must admit that this absence of contradiction to experiments is highly problematic. This is because all the new modes are strongly coupled which makes the analysis unreliable. Moreover, it is not simply about a breakdown of naive perturbation theory. It is an infinitely strong coupling making the very initial data problem ill-posed.

Given these side effects of the random zoo of remnant symmetries, one is tempted to generalise it even further.  One possible choice would be a model of $f({\mathbb T},{\mathbb B})$ type. Those go beyond one of the main initial motivations for $f({\mathbb T})$ gravity, for they do produce 4-th order equations of motion. It is unclear whether they can avoid the Ostrogradski-type ghosts, unless in the case of $f(\mathop{R}\limits^{(0)})$. However, what is absolutely clear is that they inherit all the troubles of $f({\mathbb T})$ gravity. Indeed, they obviously can be rewritten as $f({\mathbb T},\mathop{R}\limits^{(0)})$, with all the same issues of severely unstable amounts of symmetries available in the system.

At the same time, I must admit that we still do not have a clear view over the whole landscape of teleparallel options. Given all the recent activity in the $f(\mathbb T)$ models, it looks very strange that, even being quite an old idea \cite{HaSh}, other quadratic invariants of the torsion tensor are much less studied up to now. In my opnion, they may be capable of providing us with interesting and more reliable options \cite{menew}.

\subsection{Symmetric teleparallel}

Having got the troubles of metric teleparallel theories, one might think of using symmetric teleparallel ones. They are almost in their infancy yet. However, the construction actually goes in a very similar fashion, and most probably shares all the deep foundational issues of the metric teleparallel approach. Let me just briefly summarise the main ideas.

If a flat connection possesses only non-metricity but no torsion, then it can be put to zero\footnote{In the covariant language, it is called the coincident gauge.} and the formula (\ref{gencon}) reduces to
$$0=\Gamma^{\alpha}_{\mu\nu}=\mathop{\Gamma}\limits^{(0)} {}^{\alpha}_{\mu\nu}+L^{\alpha}_{\hphantom{\alpha}\mu\nu}$$
with the disformation tensor being
\begin{equation}
\label{disf}
L_{\alpha\mu\nu}=\frac12 \left(Q_{\alpha\mu\nu}-Q_{\mu\alpha\nu}-Q_{\nu\alpha\mu}\right)
\end{equation}
in terms of  the non-metricity tensor $Q_{\alpha\mu\nu}= \partial_{\alpha} g_{\mu\nu}$.

All the rest goes in just the same way as for the metric teleparallel cases above. One easily relates the two curvature tensors to each other and finds that
\begin{equation}
\label{basrel2}
0={\mathbb R}=\mathop{\mathbb R}\limits^{(0)}+\mathbb Q + \tilde{\mathbb B}
\end{equation}
with the non-metricity scalar
\begin{equation}
\label{nonmetsc}
{\mathbb Q}=\frac14 Q_{\alpha\mu\nu}Q^{\alpha\mu\nu}-\frac12 Q_{\alpha\mu\nu}Q^{\mu\alpha\nu}-\frac14 Q_{\mu}Q^{\mu}+\frac12 Q_{\mu}\tilde Q^{\mu}
\end{equation}
and the new boundary term
$$\tilde{\mathbb B}=g^{\mu\nu}\mathop{\bigtriangledown}\limits^{(0)}{}_{\alpha}L^{\alpha}_{\hphantom{\alpha}\mu\nu}-\mathop{\bigtriangledown}\limits^{(0)}{}^{\beta}L^{\alpha}_{\hphantom{\alpha}\alpha\beta}=\mathop{\bigtriangledown}\limits^{(0)}{}_{\alpha}\left(Q^{\alpha}-\tilde Q^{\alpha}\right)$$
where the traces are defined as
$$Q_{\alpha}\equiv Q^{\hphantom{\alpha}\mu}_{\alpha\hphantom{\mu}\mu}\qquad  \mathrm{and} \qquad \tilde Q_{\alpha}\equiv Q^{\mu}_{\hphantom{\mu}\mu\alpha}.$$

I guess, there is no need for repeating the very same story  for the scalar $\mathbb Q$ which has been told in the previous subsection about the scalar $\mathbb T$. When going from the new GR-equivalent model (STEGR) to, say, an $f(\mathbb Q)$ theory, the diffeomorphisms get broken, similar to the fate of Lorentz symmetries before. Of course, it produces new degrees of freedom. But intuition tells me that it can hardly go in a stably broken way either. At least for the Minkowski background, the strong coupling is quite obvious again.

\section{The meaning of the zero-spin-connection tetrad}

Note that in the previous section I have fixed the Weitzenb{\"o}ck gauge and the coincident gauge for the metric version and the symmetric version of teleparallel theories respectively. Actually, both approaches can be taken as going in the gauge of vanishing spin-connection for a tetrad \cite{mesym} defining the Weitzenb{\"o}ck connection (\ref{Weitz}).

It is obvious for the metric teleparallel connection, but is also true of the symmetric teleparallel case. Indeed, from the vanishing connection $\Gamma^{\alpha}_{\mu\nu}=0$ by a coordinate change $x\longrightarrow \xi(x)$ one can get\footnote{Note the second derivatives in the action then!}
\begin{equation}
\label{gensym}
\Gamma^{\alpha}_{\mu\nu}=\left[(\partial\xi)^{-1}\right]^{\alpha}_{\beta} \partial_{\mu} \partial_{\nu} \xi^{\beta}.
\end{equation}
Basically, the functions $\xi^n$-s are a set of coordinates in which the spacetime connection is zero. The formula (\ref{gensym}) obviously means that the symmetric teleparallel geometry can be described in terms of a set of 1-forms that form a covariantly-constant basis
$$e^n_{\mu}\equiv\frac{\partial\xi^n}{\partial x^{\mu}},$$
or a (co-)tetrad with zero spin-connection, and the affine connection coefficients (\ref{Weitz}). This is a basis of coordinate vectors, with the coordinates $\xi^n$ in the role of the Cartesian ones.

We can also approach this result from a more general track. Suppose a connection $\Gamma^{\alpha}_{\mu\nu}$ has zero curvature tensor. It means that the parallel transport is unambiguously defined. Once I've chosen a vector, or a basis of vectors $e^{\mu}_n$, at a given point, I can parallelly transport it to everywhere else, and the result does not depend on the path taken, modulo possible global effects. In other words, I've got a field of covariantly constant vectors which can be used as a zero-spin-connection tetrad.

From the defining equality\footnote{It is taken as just for a collection of 1-forms, therefore no spin connection is assumed there.} $\bigtriangledown_{\mu}e^n_{\nu}=0$, we can easily derive
\begin{equation*}
\Gamma^{\alpha}_{\mu\nu}=e_n^{\alpha}\partial_{\mu}e^n_{\nu},
\end{equation*}
or the definition (\ref{Weitz}) again, now for a tetrad of an arbitrary type. At least locally, every flat connection can be written like this, for some particular tetrad. In other words, the zero-spin-connection tetrad has a clear geometrical meaning: it is a covariantly constant basis of vector fields.

Let me summarise it once more. A geometry being teleparallel, i.e. of zero curvature, means that there exists a basis of covariantly conserved vectors, or its dual basis of  covariantly constant 1-forms, $\nabla_{\mu}e^a_{\nu}=0$,  or equivalently a soldering form which corresponds to zero spin connection. It is true of any teleparallel model. The two basic variants of these models are obtained by the following specifications:

In {\it metric teleparallel}, the usual approach is that this tetrad as a dynamical variable is absolutely free (for sure, apart from non-degeneracy), while the metric is defined as $$g_{\mu\nu}=\eta_{ab} e^a_{\mu} e^b_{\nu},$$ so that an arbitrary tetrad is orthonormal by definition. 

In {\it symmetric teleparallel}, the tetrad is holonomic, i.e. it is a basis of coordinate vectors $$e^a_{\mu}=\frac{\partial\xi^a}{\partial x^{\mu}},$$ while the metric is an independent variable.

I would like to stress that, due to its very geometric meaning, {\it the teleparallel connection should not be invariant under local transformations of its defining tetrad.} This tetrad defines the notion of the flat parallel transport, and it is an objective feature of a given geometry, not something to be freely chosen. Of course, if needed or desired, one can always rewrite all the theory in terms of another tetrad, precisely like one can use any curvilinear coordinates in a Euclidean space. However, the fundamental teleparallel tetrad used in the connection (\ref{Weitz}) is its objective defining feature, the same way as the Cartesian coordinates are very important and objectively existing geometric objects in the Euclidean geometry.

\section{Discussion}

In  general,  I don't think that phenomenologically it is a reasonable idea, to go for teleparallel formulations of gravity. The observed geometry of this world is anyway of Riemannian nature. Looking at the motion of test particles, we basically observe the geodesics of the standard pseudo-Riemannian geometry, with the metric structure which is also kind of given to us by the rulers and clocks in our hands. It is not a full-fledged rigorous derivation of its reality. However, strictly speaking, I cannot even be sure that the audience I am talking to does objectively exist, beyond my own imagination. But to all the reasonable standards of reliability, the standard geometry of GR is indeed confirmed by what we see around.

Of course, we can and should always be ready for admitting some deviations from the theories we are used to, be it torsion or non-metricity, or extra dimensions, or something totally unexpected. At the same time, the teleparallel approach does not simply add a new entity as a small correction. It totally changes the foundations. And then we tend to assume a deep physical meaning in a very esoteric quantity. How do we ever observe the flat parallel transport of teleparallel gravity? As long as we don't go away from the GR-equivalent models, there is simply no way of doing so, for those have the equations of motion invariant under any change in the choice of the teleparallel structure.

In the last years, a statement can often be seen that these approaches have finally solved the problem of energy \cite{energy}. However, it's not even possible to objectively define what was the problem and what is this energy which those people so much dream of having. If we take it as an integral of motion related to symmetry under shifts in time, then it simply does not apply to gravity for non-existence of any objective time. If we take it as just a tool for analysing equations, then there is no correct or incorrect choice of an integral of motion. Just anything goes, as long as there is no mistake in calculations. Then justfying any particular quantity, which people may like or dislike, by some esoteric and unobservable generalised geometric entity simply makes no sense \cite{noener}.

I agree that here we run into problems with the usual understanding of quantum physics. However, it is just because these two theories are absolutely incompatible, to start with. And the problem is not about some technicalities. I would say that, most importantly, it is the question of time. Quantum physics cannot exist without it. And then even the special relativity already had led it to the quantum field theory which still lacks a mathematically rigorous foundation. 

Of course, we can still keep going with our eyes closed. However, the amazing fact is that the two main parts of modern physics do teach us absolutely different and contradictory attitudes to the notion of time. And if there is any good logic to be behind the structure of the physical world, this situation cannot continue without a limit. In my opinion, we do not understand something very important about our universe. Some parts of our knowledge are definitely to be considerably changed in the future, and personally I would rather go with gravity which is mathematically nice.

To be clear, I do not claim having ultimate answers to anything. However, recalling also the measurement problem, and intentionally putting some part of a joke into the phrase, I would say that quantum physics has got nothing reasonable in it, except for the mere fact that it perfectly corresponds to all experimental data. In a sense, it looks like some effective description of something we don't really understand or have control of, like in the case of heat transfer which also does see some "objective" time and gets the first time-derivative order in the equation.

All in all, I don't observe anything to tell us that the teleparallel ideas are the next step to be taken by gravitational science. Moreover, all the simplest teleparallel modifications do not seem to be viable at all \cite{issues}, not even from purely theoretical perspective. Notwithstanding all that, I think that studying them is very interesting and important, also for possibly getting a better understanding of General Relativity itself.

{\bf A brief remark on another opinion.} After this contribution to the conference proceedings had been finalised, an interesting paper \cite{newcov} with the covariant viewpoint appeared. It claims that the covariant approach is the only way of achieving mathematical consistency of the teleparallel framework because "the change of a basis is a fundamental concept connected with the very definition of the manifold". This is of course wrong. Mathematically, neither coordinates nor any other bases are needed for formal treatmeant of differential manifolds. However, the point is that, if we do introduce an arbitrary tangent space basis and want to keep all the previously defined quantities unchanged when decomposing over it, that would indeed require the notion of spin connection. I have no objection to that. However, every teleparallel connection is uniquely related to a preferred basis, up to a global transformation. As long as we take this basis not as an arbitrary one but as a fixed full linearly-independent set of vector fields, the pure-tetrad-approach torsion is a genuine tensor. Moreover, this is a very mathematical view. Namely, in absolute teleparallelism, the tangent bundle gets identified with a trivial bundle. Needless to say, if one wants to use an arbitrary basis in it, one can do  so by having two different bases there: an arbitrary one and the preferred one. The usual covariant approach \cite{newcov} is the very same thing in a more contrived realisation of using an arbitrary tetrad and a linear transformation to the preferred one, instead of that basis itself.

\end{document}